# Application of the first-passage time method to tribological problems

V. V. Ryazanov

Institute for Nuclear Research, pr. Nauki, 47 Kiev, Ukraine, e-mail: vryazan19@gmail.com

Highlights

All real physical (tribological) processes occur with changes in entropy.
We assume that the first-passage time is equal to the time to failure.
Method the first-passage time can be effective in tribological problems.

The time until the failure of some node of the system or until the end of some stage of the operation of the tribological system is associated with the change in entropy in the system that occurs during this time. Methods of the first-passage time by a random process of some given level are used. We assume that the first-passage time is equal to the time to failure. A statistical distribution containing the first-passage time is introduced. The thermodynamic parameter associated with the random variable of the first-passage time is expressed in terms of the change in entropy. This parameter is also included in the expression for the time to failure. This approach allows arbitrary reasons for failures to be included in the consideration. The possibilities of the proposed approach are discussed.

## 1. Introduction

In works [1-17] the relations of entropy with problems of tribology are considered. In [13-17], it is shown how methods of statistical physics and probability theory are applied to tribological problems. However, the first-passage time (FPT) methods [18-21], which are effective in application to problems of the time to reach failure or the end of a certain stage of the process, have not been applied to tribological studies. For example, definitions of time to failure in tribology are directly related to FPT. The aim of this work is to apply the FPT methods to tribological problems. The methods proposed in the paper are based on the work [20]. In real physical systems, changes in entropy are constantly occurring. In this paper, we consider the effect of entropy changes on the time to failure caused by tribological reasons.

In work [20], the first-passage time (FPT) is associated with the entropy changes in the system that occur during this time. The impact of entropy changes on the mean FPT can be significant [21]. Such estimates can be made for the time until the failure of some node of the system or until the end of some stage of the operation process. Data on entropy changes are taken from tribological studies. The key issue is the choice of distribution for the FPT. One of the possible solutions can be the maximum entropy method with restrictions corresponding to the specifics of the process. This approach makes it possible to isolate the contributions from various processes (friction, wear, structural-phase transformations, physical-chemical processes, etc.), evaluate them, and choose the optimal ways of influencing the system.

We assume that the first-passage time is equal to the time to failure. What is first-passage time (FPT)? This is the time for the random process to reach some limit.

The purpose of this paper is to show the possibilities of using the FPT method in tribological problems. The main advantage of the approach proposed in [20] is the relationship between FPT and entropy changes over the FPT period. Entropy changes are often determined in tribological studies.

In the second section, the general scheme for taking into account entropy effects in the FPT method is considered. The third section considers the possibility of determining the thermodynamic parameter conjugated with FPT, which contains the entropy changes during the FPT. It is noted that a similar value is present in tribology. In the fourth section, a specific stochastic process model is selected, for which explicit expressions are written that characterize the achievement of a given level. In the fifth section, brief conclusions are given.

## 2. Statistical description of the expected service life

In [20–24], a generalization of the Gibbs distribution to the nonequilibrium case is introduced, containing FPT as a thermodynamic parameter. The introduction of such a distribution is based on the theory of large deviations and the maximum entropy method. In tribology, in addition to a large number of particles in a system, there are a large number of influences. For example, friction is accompanied by multiple influences. Both the initial size of tribological elements and their wear rates depend on a large number of independent factors. In this case, a statistical distribution is used, which contains the time it takes for a random process to reach a certain level (first-passage time [22, 23] or lifetime in terms of [24]).

In [20, 21, 22–24], the random value of the first passage time (lifetime) $T_\gamma$ of a random process was considered as a thermodynamic parameter. A statistical distribution with this parameter was introduced. The microscopic density of this distribution in the phase space of variables $z$ (coordinates and momenta of all particles of the system) has the form

$$\rho(z;u,T_\gamma) = \exp\{-\beta u - \gamma T_\gamma\}/Z(\beta,\gamma) \qquad \beta = 1/T_1, \qquad (1)$$

where $u$ is the energy of the system (considered as a random process), $T_1$ is the temperature (in energy units),

$$Z(\beta,\gamma) = \int \exp\{-\beta u - \gamma T_\gamma\}dz = \int\int du\, dT_\gamma\, \omega(u,T_\gamma)\exp\{-\beta u - \gamma T_\gamma\}, \qquad \beta = 1/T_1, \qquad (2)$$

is the partition function. Equality in expression (2) corresponds to the transition from a microscopic description to a macroscopic one. The Lagrange parameters $\beta$ and $\gamma$ are determined from the equations for the averages:

$$\langle u \rangle = -\partial \ln Z / \partial \beta \big|_\gamma; \qquad \langle T_\gamma \rangle = -\partial \ln Z / \partial \gamma \big|_\beta \quad <u> = -\partial \ln Z/\partial \beta \big|_\gamma; \qquad \beta = 1/T_1. \qquad (3)$$

In [20–24] cells of the "extended" (compared to the Gibbsian) phase space with constant values ($u$, $T_\gamma$) are introduced (instead of phase cells with constant values of $u$). The structure factor $\omega(u)$ is replaced by $\omega(u,T_\gamma)$ is the volume of the hypersurface in the phase space containing fixed values of u and $T_\gamma$. In this case $\int\omega(u,T_\gamma)dT_\gamma = \omega(u)$. The function $\omega(u,T_\gamma)$ corresponds to the internal properties of the system. At the mesoscopic level, this function is described by a random process, and $\omega(u,T_\gamma)$ is the joint probability density of unperturbed (for $\gamma=0$ and $\beta = \beta_{eq}$ is an equilibrium state (or some stationary nonequilibrium) and without external influences) values $u$, $T_\gamma$ understood as stationary distribution of this process. Distribution (1)-(2) generalizes the Gibbs distribution, in which $\gamma=0$, and is valid for stationary non-equilibrium systems. In [25] it is shown that non-stationary processes can also be described by this kind of distribution.

The joint distribution density of random variables $u$ and $T_\gamma$ is equal to

$$p(u,T_\gamma) = \int \delta(u - u(z))\delta(T_\gamma - T_\gamma(z))\rho(z;u(z),T_\gamma(z))dz =$$

$$= \rho(u,T_\gamma)\omega(u,T_\gamma) = \exp\{-\theta^{-1}u - \gamma T_\gamma\}\omega(u,T_\gamma)/Z(\beta,\gamma) \tag{4}$$

Integrating (4) over $T_\gamma$, we obtain a distribution for the energy $u$ of the form

$$p(u) = \int p(u,T_\gamma)dT_\gamma = \frac{e^{-u\beta}}{Z(\beta,\gamma)}\int_0^\infty \omega(u,T_\gamma)e^{-\gamma T_\gamma}dT_\gamma, \qquad \beta = 1/T_1. \tag{5}$$

In accordance with the assumptions about the form of the function $\omega(u,T_\gamma)$, we rewrite it in the form

$$\omega(u,T_\gamma) = \omega(u)\omega_1(u,T_\gamma) = \omega(u)\sum_{i=0}^n P_i f_i(T_\gamma,u). \tag{6}$$

Expression (6) assumes that the system has $n+1$ classes of ergodic states, $P_i$ is the probability that the system is in the $i$-th class of ergodic states, $f_i(T_\gamma,u)$ is the probability density that in a system that is in ergodic states of the $i$-th class, the time of the first passage is $T_\gamma$. The expression under the integral on the right side of (5) determines the non-equilibrium part of the distribution.

If we choose the function $f_i$ in (6) in the form of a gamma distribution

$$f_i(x) = \frac{1}{\Gamma(\alpha_i)}\frac{1}{b_i^{\alpha_i}}x^{\alpha_i-1}e^{-x/b_i}, \quad x>0, \quad f_i(x)=0; \quad x<0; \quad \int_0^\infty e^{-\gamma_i x}f_i(x)dx = (1+\gamma_i b_i)^{-\alpha_i}, \tag{7}$$

where $\Gamma(\alpha)$ is the gamma function, and put $b_i\alpha_i = T_{\gamma 0i}$, ($T_{\gamma 0i}$ are the unperturbed average lifetimes of the system in $i$-th states [20-24]), $\alpha_i = \gamma_i/\lambda_i$, where $\lambda_i$ is the intensity of entry into the $i$-th subsystem in a state of dynamic equilibrium, $\gamma_i = \gamma$ in the $i$-th subsystem, then from (4)-(7) we obtain that

$$(1+\gamma_i b_i)^{-\alpha_i} = (1+\lambda_i T_{\gamma 0i})^{-\gamma_i/\lambda_i}; \quad p(u) = \int p(u,T_\gamma)dT_\gamma = \frac{e^{-\beta u}\omega(u)}{Z(\beta,\gamma)}\sum_{i=0}^n P_i/(1+\lambda_i T_{\gamma 0i})^{\alpha_i}. \tag{8}$$

If $f_i$ is an exponential distribution, then $\alpha_i=1$, $\gamma_i=\lambda_i$, $b_i=T_{\gamma 0i}$.

For the exponential distribution when $\alpha_i=1$ in (7)-(8) and one class of ergodic states when $n=0$,

$$(1+\gamma T_{\gamma 0})^{-1} = <T_\gamma>/T_{\gamma 0}, \tag{9}$$

where $<T_\gamma>$ is the average lifetime of the system obtained from (2)-(3), (6)-(7) [20-24].

Let us define partition function (2). In this case, we use expressions (2)-(3) and the approximation that the distribution of the first passage time does not depend on the random value of energy, the variables are separated, and [20]

$$Z(\beta,\gamma) = Z_\beta Z_\gamma, \quad Z_\beta = \int e^{-\beta u}\omega(u)du \quad Z_\gamma = \int_0^\infty e^{-\gamma T_\gamma}f(T_\gamma)dT_\gamma, \qquad \beta = 1/T_1, \tag{10}$$

$$Z_\gamma = \int_0^\infty e^{-\gamma T_\gamma}\sum_{j=0}^n P_j f_j(T_\gamma,\Phi)dT_\gamma, \tag{11}$$

where $f(T_\gamma)$ is the probability density of the distribution of the *FPT*.

In (10), we assume that the parameters included in the distribution of the first passage time depend on the average values of the energy and flux, and not on their random values. In accordance with expressions (10)-(11), the partition function factor describing the non-equilibrium behavior of the system is expressed through the Laplace transform of the probability density the first passage time is $T_\gamma$.

Besides distribution (7), many other distributions for FPT can be used. It depends on the specific task.

Assuming that *FPT* can be instrumentally measured, we introduce the local specific entropy $s$ corresponding to the distribution (1) ($u$ is specific internal energy) by the relation [20]

$$s = -\langle \ln \rho_{rel}(z;u,T_\gamma) \rangle = \beta \langle u \rangle + \gamma \langle T_\gamma \rangle + \ln Z(\beta,\gamma); \qquad ds = \beta d\langle u \rangle + \gamma d\langle T_\gamma \rangle. \qquad (12)$$

We rewrite (12) in the form of the relationship of expressions for equilibrium and non-equilibrium entropy. The parameter $\gamma$ of distribution (1) will be related to entropy using the generally accepted definition of entropy in statistical physics as the logarithm of the distribution density (1) averaged over this distribution, $s = \langle \ln \rho(z,u,T_\gamma) \rangle$ (12), where brackets denote averaging. Variables are separated as in (10). In the case of one class of ergodic states, from (6) we obtain $\omega(u,T_\gamma) = \omega(u) f(T_\gamma)$, $\omega(T_\gamma) = f(T_\gamma)$. The distribution density (4) is equal to $p(u,T_\gamma) = \dfrac{e^{-\beta u} \omega(u)}{Z_\beta} \dfrac{e^{-\gamma T_\gamma} f(T_\gamma)}{Z_\gamma} = p(u) p(T_\gamma)$, where $Z_\beta$, $Z_\gamma$ defined in (10). The entropy of this distribution is $s = -k_B \int p(u,T_\gamma) \ln[p(u,T_\gamma)] du dT_\gamma = s_\beta + s_\gamma$, $s_\beta = \beta \bar{u} + \ln Z_\beta$, $s_\gamma = \gamma T_\gamma + \ln Z_\gamma$. We took into account that for the nonequilibrium case in [25], the relation $s = s_B + \langle s(B) \rangle$, $s_B = -k_B \int p(B) \ln[p(B)] dB$, $\langle s(B) \rangle = k_B \int p(B) \ln[\omega(B)] dB$ was obtained; $B$ are random internal thermodynamic parameters, functions of dynamic variables $z$; in our case $B_1 = u$, $B_2 = T_\gamma$. The total uncertainty in the system is equal to the sum of the uncertainty of the parameters $B$ and the average uncertainty of the dynamic variables remaining after fixing the parameter $B$. Expanding in a power series in powers $\gamma$ value $s_\gamma$, we get $s_\gamma = -\gamma^2(\langle T_0^2 \rangle - \langle T_0 \rangle^2) \leq 0$, $s \to s/k_B$, entropy is divided by $k_B$, Boltzmann's constant.

Thus, the terms $\int \dfrac{e^{-\beta u} \omega(u)}{Z_\beta} \ln[\omega(u)] du$, $\int \dfrac{e^{-\gamma T_\gamma} f(T_\gamma)}{Z_\gamma} \ln[f(T_\gamma)] dT_\gamma$ cancel and

$$s = s_\gamma + s_\beta = s_\beta - \Delta = \gamma \bar{T}_\gamma + \beta \bar{u} + \ln Z = \beta \bar{u} + \ln Z_\beta - \Delta, \quad -\Delta = s_\gamma = s - s_\beta, \quad -\Delta = s_\gamma = \gamma \bar{T}_\gamma + \ln Z_\gamma, \qquad (13)$$

where the quantities $Z_\gamma$, $\bar{T}_\gamma$, are defined in the relations (2)-(3), (10), $\Delta \geq 0$; $Z_\beta$ is partition function; $\bar{u}$ is energy; $s_\beta$ is entropy related to the parameter $u$, ($s$ is the entropy density). We got a match with the expression $s = \langle \ln \rho(z,u,T_\gamma) \rangle$. Stationary part of entropy is $s_{st} = s_\beta = \beta u_\beta + \ln Z_\beta$. The Shannon entropy $s_\gamma = -\int P(\Gamma) \ln P(\Gamma) d\Gamma$ is equal $s_\gamma = \gamma \bar{T}_\gamma + \ln Z_\gamma + s^\beta{}_\gamma$, where

$$s^\beta{}_\gamma = -\int \dfrac{e^{-\gamma T_\gamma}}{Z_\gamma} [f(T_\gamma)] \ln[f(T_\gamma)] dT_\gamma. \qquad (14)$$

Taking into account (13), we obtain that $\langle s(B) \rangle = k_B \int p(B) \ln[\omega(B)] dB = -s^\beta{}_\gamma$, and expression $s_\gamma = -\int P(\Gamma) \ln P(\Gamma) d\Gamma$ leads to (13).

### 3. Determination of the conjugate FPT thermodynamic parameter γ.

The thermodynamic parameter $\gamma$ of distribution (1), expressed in terms of entropy changes, can be determined in various ways. After substituting the parameter $\gamma(\Delta)$ into the expression for the average $T_\gamma$ (for example, (26)) we get the dependence $T_\gamma$ on of entropy changes. After substituting the parameter $\gamma(\Delta)$ into the expression for the average $T_\gamma$ (for example, (26)) we obtain the dependence on the change in entropy. In this section, we will indicate three such methods. Let's start with a direct analogy between the parameter $\gamma$ and the wear parameter in tribology. In [16, 17], the controlled parameter of the state of the tribological system $X_t$, which changes with wear, the linear size of one of the parts or their combination, is assumed to be a random variable. In the linear wear model, the state parameter $X_t$ changes as

$$X_t = X_0 - \Gamma_y t, \qquad (15)$$

where $X_0$ is random variable of the state parameter $X_t$ at the initial time $t_0$, $\Gamma_y$ is a random variable linear wear rate with mean value $\gamma_y$. The limiting (normative) value of the state parameter $X_t$ is equal to $X_{lim}$. In [16], the parameter $\gamma_y$ is expressed in terms of the ratio of the nominal areas of contact and friction of the elements, the critical value of the measure of damage to materials, the speed of the relative movement of the surfaces, and the coefficients of conversion (absorption) of external energy by the surface layer of triboelements. In [17], this parameter is expressed through the volumetric wear of each interface element on the friction path and the friction area of the elements. In [17], the degradation or linear wear rate parameter is defined as

$$\gamma_y = \nu p_{max} f_{mech} V_{sl} / \Delta u_c^*. \qquad (16)$$

Here $\nu$ is the coefficient of absorption of external energy by the surface layer of the sealing material, determined as a function of the physical and microgeometric characteristics of the friction pair elements [17]; $p_{max}$ is the maximum nominal pressure at the contact of the plunger and the seal during operation, determined using a known technique for interference mates; $f_{mech}$ is the mechanical component of the friction coefficient; $V_{sl}$ - speed of relative sliding of elements; $\Delta u_c^*$ is critical density of internal potential energy (critical energy capacity) of the material of the contact volumes of the seal. The critical energy intensity is defined as [17]: $\Delta u_c^* = \Delta H_S - \Delta H_T$. Here $\Delta H_S$ is the enthalpy of melting of the sealing material in the liquid state at the melting temperature $T_S$; $\Delta H_T = \int_0^T \rho c \, dT$ are enthalpy of the material at temperature $T$ of the steady friction mode, which, in turn, is determined by known methods; $\rho$, $c$ – density and heat capacity of the sealing material at temperature $T$. The numerical characteristics of the random variables $\Gamma_y$ included in the conditions (15) are determined by equation (16) as a function of the numerical characteristics of the mechanical component $f_{mech}$ of the friction coefficient. They, in turn, are determined as a function of the numerical characteristics of the complex index of plunger surface roughness over the range of reference data for run-in surfaces according to the "three sigma" rule [26].

The second approach to determining this quantity is related to the expression for the distribution entropy for *FPT*, as in [27] or as in estimating the rates of entropy production [29-31].

In the third approach, as in [20], an explicit expression for the change in entropy is used, which is assumed to be known, and the equation (13).

In [20], the equation for determining the parameter was written in terms of the change in entropy. For example, for the case of a phase transition, the change in entropy is

$$\Delta S = \frac{\Delta H}{T_{phase}}. \qquad (17)$$

Phase transitions are associated with surface melting and recrystallization of metals, occurring with a change in entropy. Here $\Delta H$ is the change in enthalpy, the latent heat absorbed or shed during the phase transition, and $T_{phase}$ is the temperature associated with the phase transition. In [28], the entropy change is written in a more general form

$$\Delta S = \Lambda_{D,PH} [(\frac{U_c}{T_m} \Delta V + \frac{G - 2\gamma_0}{T_{cr}})\Lambda_D + \frac{\Delta H}{T_{phase}}], \qquad (18)$$

where $\Lambda_D$ is the function (coefficient) of the interaction between the entropy components during surface plastic deformation and cracking, $\Lambda_{D\backslash PH}$ is the function (coefficient) of the interaction between the entropy components during mechanical failure and changes in material properties during the phase transition. Plastic deformation associated with abrasive wear, frictional ploughing, and/or cutting, with a change in entropy. Here $U_c$ is the work (plastic deformation or cutting) spent per unit of affected volume $\Delta V$, and $T_m$ is the temperature of the acting material medium. Fracture is associated with fatigue wear and surface damage, with a change in entropy. Here $a$ is the crack length, $G = -\partial U_S/\partial a$ is the energy release rate depending on the strain energy $U_S$, $\gamma_o$ is the surface energy, and $T_{cr}$ is the temperature of the cracked material at the crack tip.

The equation for determining the parameter $\gamma$ is associated with equation (13) or with the expression

$$-\Delta = s_\gamma = \gamma \overline{T}_\gamma + \ln Z_\gamma. \tag{19}$$

If expression (18) is substituted into (19) we obtain an equation for $\gamma$.

To determine the parameter $\Delta$, methods for estimating the rates of entropy production are applicable [29-31]. Estimator of dissipation that we review is the Thermodynamic Uncertainty Ratio (*TUR*) [29, 30]. The *TUR* has been proposed as an estimator of dissipation as it lower bounds the entropy production rate in stationary Markov jump processes and overdamped Langevin processes. In this case $\Delta_{TUR} \approx \dfrac{2\Phi^2}{D_\Phi}$, where $D_\Phi = \langle \Phi^2 \rangle - \langle \Phi \rangle^2$ is flux $\Phi$ dispersion. If used to determine the average flux $\Phi = \overline{\Phi}$ and flux dispersion $D_\Phi$ Maxwell distribution under the assumption of a constant density and fluctuations in velocity, then; $\Delta \to \Delta s / k_B$, $k_B$ is Boltzmann's constant; $\Delta \approx 4.41$. Similar results are obtained by using an estimator based on the first-passage ratio (*FPR*) [31].

Let us find the parameter $\gamma$ from relation (19). Expanding the right side of (19) in powers of $\gamma$ up to the second order, we obtain the relation

$$-\Delta \approx \frac{1}{2}\gamma^2 \frac{\partial \overline{T}_\gamma}{\partial \gamma}\bigg|_{\gamma=0}, \quad \gamma \approx [-2\Delta / \frac{\partial \overline{T}_\gamma}{\partial \gamma}\bigg|_{\gamma=0}]^{1/2}. \tag{20}$$

## 4. Stochastic process model

It is required to establish an adequate correspondence between a physical phenomenon and a random process used for its mathematical modeling. There are many random processes that can be adapted to a specific task. The properties of the process can be significantly affected by various circumstances, for example, changes in the boundary conditions [32, 33].

The same phenomenon under different conditions can exhibit different properties and be described by different random processes, distributions and their Laplace transforms. Examples of this kind are given in [34]. For our purposes, the Laplace transforms of the FPT distributions are more important than the distributions themselves. This facilitates the analytical solution of the problem.

The phenomenon under study is modeled by some random process. The functional of this random process is the FPT statistics. The exact analytical form of the solution for the FPT is determined in rare cases. It is possible to use general results for the FPT, for example, the first few terms of the expansions obtained in [35-36] for the Laplace transform of the FPT. There are other options for selecting and configuring FPT statistics models. The use of various types of approximations also significantly depends on the stage of system evolution [37, 38].

The proposed approach is based on probabilistic mathematical relations and has great generality. The area of applicability of the results is large and, possibly, corresponds to the area of applicability of the Gibbs statistics. This generality allows one to consider arbitrary distributions rather than being restricted to any one particular form of FPT.

Let us illustrate the application of the above approach by examples. Let's apply the above possibilities of linking the FPT distribution with the thermodynamic characteristics of the system to the linear wear model (15). Consider reaching the limit value of the parameter $X$ (15) $X_{lim}$ as a random process reaching a certain level. As an independent thermodynamic parameter, a random time of the first achievement of the set value $X=X_{lim}$ is chosen. Adding this parameter, along with the energy, to the exponential distribution, as in (8), we write the relation for the entropy of the form (12), (14) taking into account (10), (12), (13).

The behavior of the collapsing segment $X_0$-$X_t$ from expression (15) in [16, 17] is modeled by the normal distribution. Let us model it with a random process of the form

$$\frac{dv}{dt} = \mu + \xi(t), \qquad (21)$$

where $v = X_0$-$X_t$ from (15) is the length of the section under consideration, $\mu$ is the rate of its destruction; in accordance with (15) $\mu = \gamma_y$ from (15), $\xi(t)$ is normal white noise; $\langle \xi(t) \rangle = 0$, $\langle \xi(t_1)\xi(t_2) \rangle = \sigma^2 \delta(t_1 - t_2)$. Process (21) is considered on the interval (0, h), where at point 0, corresponding to point $v = X_0$-$X_t=0$ from (15), a reflecting screen is placed, and at point h, corresponding to point $X_{lim}$ from (15), absorbing screen. In [32], a differential equation was obtained for the probability density of the first-passage time reaching the boundary h. After applying the Laplace transform to this equation, taking into account the initial condition, a second-order ordinary differential equation was obtained in [35] for the Laplace transform of the probability density of the first-passage time reaching the boundary h. In our case, this is the value $Z_\gamma$ (11). The general solution of this equation is given by

$$Z_\gamma = Z_{(0,h)}(s = \gamma, v_0) = A e^{\gamma_1 v_0} + B e^{\gamma_2 v_0}, \qquad (22)$$

where $s=\gamma$ from (11), $v_0$ is initial value of random variable $v$; accept $v_0=0$, constants $A$ and $B$ are determined from the boundary conditions of the

$$A e^{\gamma_1 h} + B e^{\gamma_2 h} = 1, \quad \gamma_1 A e^{\gamma_1 c} + \gamma_2 B e^{\gamma_2 c} = 0, \quad c = 0, \qquad (23)$$

and $\gamma_1, \gamma_2$ are roots of the characteristic equation

$$\gamma_{1,2} = \frac{-\mu \mp \sqrt{\mu^2 + 2\sigma^2 s}}{\sigma^2}. \qquad (24)$$

From the boundary conditions (23) we find the constants $A$ and $B$ and for the Laplace transform (11) we obtain the expression

$$Z_{s(=\gamma)} = \frac{\gamma_2 - \gamma_1}{F}, \quad F = \gamma_2 e^{\gamma_1 h} - \gamma_1 e^{\gamma_2 h}, \qquad (25)$$

where the numerator and denominator in (25) are positive at $s>0$, $h>0$; $Z_s<1$.

The average time to reach the boundary h, obtained from (25), is

$$\bar{T}_s = -\frac{\partial \ln Z_s}{\partial s} = \frac{1}{\sqrt{\mu^2 + 2\sigma^2 s}} \frac{1}{F}(\frac{\partial F}{\partial s} - \frac{\sigma^2}{\sqrt{\mu^2 + 2\sigma^2 s}}), \quad \frac{\partial F}{\partial s} = (1-\gamma_2 h)e^{\gamma_1 h} + (1-\gamma_1 h)e^{\gamma_2 h}. \qquad (26)$$

In relationship (24)-(26) it is necessary to substitute the values $s=\gamma(\Delta)$ defined in Section 3. Wherein $\bar{T}_s \geq 0$, $\frac{\partial \bar{T}_s}{\partial s} \leq 0$. With an increase in the value $s=\gamma$, the average time $\bar{T}_\gamma$ to reach the boundary $h = X_{lim}$ decreases.

Note that, in addition to process (21), it is possible to use both a number of other continuous random processes and a number of discrete random processes, if we assume the discreteness of the destruction process of the section $v=X_t$-$X_0$. For example, if in (21) both boundaries 0, and $h$ are absorbing,

$$Z_\gamma = \int_0^\infty e^{-\gamma t} f(t)dt = \frac{e^{\gamma_2 h_0}(1-e^{\gamma_1 h}) - e^{\gamma_1 h_0}(1-e^{\gamma_2 h})}{(e^{\gamma_2 h} - e^{\gamma_1 h})}, \quad h_0 = h(t=0).$$

Once again, we note that for each specific task, it is necessary to look for that random process that most fully corresponds to this particular task. In [39] and [40] *FPT*s are considered for achieving a certain level of entropy production and flux values.

## 5. Conclusion

The value of the lifetime $T_{\gamma=0} = T_0$ in (26) is assumed to be known, its change is determined under the influence $\gamma(\Delta)$ of the processes occurring in the system. For distribution (21)-(26), the mean time to failure can only decrease. There are distributions for which an increase $T_0$ is possible in some intervals of entropy, for example, the Weibull distribution, Feller processes, etc.

In [38] the entropy change rate which reflects the relation between the entropy and cumulative plastic strain can be determined:

$$\dot{s} = -\frac{N_0 k_B \alpha}{m_s \ln(p_{cr}/p) f(\sigma_m/\sigma_{eq})} [\dot{f}(\sigma_m/\sigma_{eq}) \ln(p_{cr}/p) - f(\sigma_m/\sigma_{eq}) \dot{p}/p], \qquad (27)$$

where $\alpha$ is the damage exponent; $\sigma_{eq}$ is the equivalent von Mises stress, $\sigma_m$ is the hydrostatic stress or mean stress ($\sigma_m = \sigma_{kk}/3$). For uniaxial loading, the factor $f(\sigma_m/\sigma_{eq}) = 1$, cumulative plastic strain rate is $\dot{p}$, $p$ is the cumulative plastic strain. The corresponding cumulative plastic strain when $D = D_0$ and $D = D_{cr}$ are $p_{th}$ and $p_{cr}$, respectively; $D$ is the damage variable; $D_{cr}$ is the value of damage variable when final failure occurs, $N_0$ is the Avogadro constant, $m_s$ is the specific mass, $k_B$ is the Boltzmann constant. Expression (27) can also be used in determining the parameter $\gamma(\Delta)$, Section 3.

The problems of determining the parameters $\gamma$ and entropy production $\sigma_s$ are related. If the values $\sigma_s$ (or $\dot{s}$ from (27)) are known, then multiplying this value by $\bar{T}_\gamma$ and substituting into equation (19), we obtain an equation for determining the parameter $\gamma$. If the value $\Delta S$ is known, then substitution of this value into relation (19) gives an equation for the parameter $\gamma$. Instead of distribution (25), many other distributions can be used, in some situations more adequately corresponding to one or another problem being solved. For example, models of random walks, death and birth, etc. It is possible to combine different methods for determining the parameters $\gamma$ and $\sigma_s$, depending on which method is more convenient, more accurately or better corresponds to the task. Thus, the proposed approach provides considerable freedom in solving specific problems.

As in [26], using the proposed methods, it is possible to consider the reliability of plunger hydraulic cylinders of the system for balancing rolls of rolling mills according to the criterion of wear resistance of sealing elements.

The relations obtained are applied to these and many other problems of tribology. Entropy changes include all possible effects on the system. This allows the description to include the behavior and use of lubricants, corrosion-fatigue degradation mechanism, the processes of friction, wear and fatigue, the processes of sliding friction with wear under sliding and mechano-sliding fatigue, etc.

Similar methods that do not use a distribution containing FPT were used in [42] and in [43]. It should be noted the general similarity between the methods of applying FPT and tribological approaches. So, in this work, a comparison of the parameter of the rate of degradation or linear wear of triboelements with the thermodynamic parameter associated with FPT is noted and used.

It was shown in [16] that the rate of degradation of a loaded element is determined by the rate of damage to the structure of its material, i. e., the rate of accumulation of the latent energy density of defects. This value is expressed through the activation energy of the process of destruction of interatomic bonds and other parameters. These relations can be used in expressions (12)-(13).

In [16], the design evaluation of the reconstruction of the rotary drive of the kiln according to the criterion of the strength of the foundation bolts was considered; - to increase the durability of the balancing system of rolls according to the criteria of wear resistance of the sealing elements of the actuating hydraulic motors; - to extend the service life of the system for cleaning hot-rolled strips from scale according to the criteria of wear resistance of friction pairs of hydraulic distributors. Problems of this kind can be solved using the methods of this work.

One of the most common reasons for the downtime of metallurgical units is the failure of spur gears. In some cases, the failure is not due to chipping or breakage of the teeth, but to the limit of wear. If the methods of design assessment of the efficiency of the spur gears according to the criterion of contact and bending strength are widely known, then predicting their durability according to the criterion of wear resistance is difficult. The results of this work are applicable to the solution of this problem.

An example of the application of the above approach can be the use of the above considerations to the probabilistic method implemented in [26] for predicting the reliability of plunger hydraulic cylinders of the system for balancing rolls of rolling mills according to the criterion of wear resistance of sealing elements.

The equations obtained on the basis of a general methodological approach to predicting the reliability of plunger hydraulic cylinders according to the criterion of wear resistance of seals make it possible, even at the design stage, to study the process of formation of their wear failures, to track the change in the level of their reliability indicators and to evaluate the resource characteristics in the expected operating conditions. Given the change in entropy, the time to failure can be determined using the above approach.

A probabilistic description based on the method FPT can be effective in tribological problems. The use of an FPT-based approach and the relationship of FPT to entropy change provides many new and promising possibilities for the study of tribological systems.